\documentclass[prl,twocolumn,showpacs,showkeys,superscriptaddress,citeautoscript]{revtex4-1}

\usepackage{amsmath,amssymb}

\usepackage[colorlinks=true,citecolor=blue,linkcolor=blue]{hyperref}
\usepackage{physics}

\newcommand*{\setR}{\mathbb{R}}
\newcommand*{\del}{\partial}
\newcommand*{\Op}{\mathcal{O}}

\usepackage{acronym}

\usepackage[T1]{fontenc}
\usepackage[utf8]{inputenc}

\usepackage{microtype}

\begin{document}

\title{Near-Conformal Dynamics at Large Charge}

\author{Domenico Orlando}
\email{domenico.orlando@to.infn.it}
\affiliation{INFN, sezione di Torino and Arnold--Regge Center, via Pietro Giuria 1, 10125 Torino, Italy}
\affiliation{Albert Einstein Center for Fundamental Physics, Institute for Theoretical Physics, University of Bern,
Sidlerstrasse 5, \textsc{ch}-3012,Bern, Switzerland}
\author{Susanne Reffert}
\email{sreffert@itp.unibe.ch}
\affiliation{Albert Einstein Center for Fundamental Physics, Institute for Theoretical Physics, University of Bern, Sidlerstrasse 5, \textsc{ch}-3012,Bern, Switzerland}
\author{Francesco Sannino}
\email{sannino@cp3.sdu.dk}
\affiliation{CP3-Origins \& the Danish Institute for Advanced Study. University of Southern Denmark. Campusvej 55, DK-5230 Odense, Denmark}

\begin{abstract}
 We investigate four-dimensional near-conformal dynamics by means of the large-charge limit.  We first introduce and justify the formalism in which near-conformal invariance is insured by adding a dilaton and then determine the large-charge spectrum of the theory.   The dilaton can also be viewed as  the  \emph{radial} mode of the \acl{eft}. We calculate the two-point functions of charged operators. We discover that the mass of the dilaton, parametrising the near-breaking of conformal invariance,  induces a novel term that is logarithmic in the charge.  One can therefore employ the large-charge limit to explore near-conformal dynamics and determine  dilaton-related  properties. 
 \end{abstract}
\preprint{}

\maketitle

\acrodef{vev}{vacuum expectation value}
\acrodef{cft}[\textsc{cft}]{conformal field theory}
\acrodefplural{cft}[\textsc{cft}s]{conformal field theories}
\acrodef{eft}[\textsc{eft}]{effective field theory}
\acrodefplural{eft}[\textsc{eft}s]{effective field theories}
\acrodef{nlsm}[\textsc{nlsm}]{non-linear sigma model}
\acrodef{ir}[\textsc{ir}]{infrared}
\acrodef{uv}[\textsc{uv}]{ultraviolet}
\acrodef{rg}[\textsc{rg}]{renormalisation group}
\acrodef{bkt}[\textsc{bkt}]{Berezinskii--Kosterlitz--Thouless}
\acrodef{qcd}[\textsc{qcd}]{quantum chromodynamics}

\acused{cft}
Conformal field theories (\textsc{cft})s play an essential role in our understanding of critical phenomena in several dimensions~\cite{Wilson:1973jj}. Of particular relevance are quantum phase transitions in four-dimensional gauge theories which are  zero-temperature transitions from conformal to non-conformal phases. A time-honored example is the number-of-flavour-driven quantum phase transitions from an \ac{ir} fixed point to a non-conformal phase where chiral symmetry is broken~\cite{Miransky:1996pd}.  Depending on the underlying mechanism behind the loss of conformality one can envision several scenarios ranging from  a \ac{bkt}-like phase transition discovered in two dimensions~\cite{Kosterlitz:1974sm} and proposed for four dimensions in~\cite{Miransky:1984ef,Miransky:1996pd,Holdom:1988gs,Holdom:1988gr,
Cohen:1988sq,Appelquist:1996dq,Gies:2005as}  to a jumping (non-continuous) phase transition~\cite{Sannino:2012wy}. The subsequent suggestion that theories with a very small number of matter fields in higher-dimensional representations could be (near) conformal~\cite{Sannino:2004qp} culminated in the well-known conformal window phase diagram of~\cite{Dietrich:2006cm} that has served as a roadmap for lattice studies~\cite{Pica:2017gcb}.   

 In all scenarios, the spectrum is not symmetric on the two sides of the quantum phase transition.  In the non-conformal phase, we have a well-defined particle spectrum with states separated by a mass gap, and depending on whether some residual global symmetries are spontaneously broken, the spectrum will feature additional gapless states. In the conformal phase, on the other hand, conformality forbids gaps enforcing a continuum of  states.  However, one can still define quasi-particles in the conformal phase if the transition occurs in a perturbative regime of the underlying theory.
In the \ac{bkt} transition, all derivatives of the correlation length with respect to the parameter driving the transition away from the symmetric phase vanish  at the critical point; in the jumping case, there is a discontinuous transition between the conformal and non-conformal phase. These are two extreme ways to characterise the four-dimensional quantum phase transition and others can be envisioned as the supersymmetric \ac{qcd} example shows~\cite{Seiberg:1994pq,Intriligator:1995au}.  If the quantum phase transition is smooth, such  as the one due to the annihilation of an \ac{ir} and \ac{uv} fixed point, soon after the transition (annihilation of the fixed points) it is natural to define three regions: a high-energy region dominated by asymptotic freedom,  a quasi-conformal region in which the coupling(s) remain nearly constant, and a low energy one where the theory develops a mass scale.  Two \ac{rg}-invariant energy scales can be naturally defined: $\Lambda_{UV}$, separating the asymptotically free behaviour from the quasi-conformal one, and the scale $\Lambda_{IR}$ below which conformality and, depending on the theory, also certain global symmetries are lost. This behaviour is colloquially known as \emph{walking} and it has been invoked several times in the phenomenological literature for models of dynamical electroweak breaking in order to enhance the effect of bilinear fermion operators~\cite{Holdom:1988gs,Holdom:1988gr}. The amount of walking is naturally measured in terms of the  \ac{rg} invariant ratio $\Lambda_{UV}/\Lambda_{IR}$. For \ac{qcd}-like theories, this ratio is of order unity while near-conformal theories of \emph{walking} type have ideally $\Lambda_{UV}/\Lambda_{IR} \gg 1$. An equivalent way to view walking is through the emergence of two complex zeros of the beta-function in the near-conformal phase~\cite{Gorbenko:2018ncu}. Perturbative examples of near-conformal dynamics have been considered in~\cite{Grinstein:2011dq,Antipin:2011aa,Benini:2019dfy}. 

Lattice methods have been developed and proven useful to explore the non-perturbative dynamics of the infrared conformal  window of gauge-fermion theories~\cite{Pica:2017gcb} while it has been proven difficult to identify and determine the nature of the quantum phase transition per se.  

A general expectation is that for a continuous quantum phase transition, a dilaton-like mode appears in the broken phase in order to account for the approximate conformal invariance~\cite{Leung:1985sn,Bardeen:1985sm,Yamawaki:1985zg,Sannino:1999qe,Hong:2004td,Dietrich:2005jn,Appelquist:2010gy}. This  dilaton-like effective action can be implemented \`a la Coleman~\cite{Coleman:1988aos} in order to saturate the underlying trace anomaly of the theory that keesps track of the breaking of Weyl invariance. Recently 
there has been renewed interest in \acp{eft} featuring the dilaton degree of freedom~\cite{Hong:2004td,Dietrich:2005jn,Goldberger:2008zz,Appelquist:2010gy,Hashimoto:2010nw,Matsuzaki:2013eva,Golterman:2016lsd,Hansen:2016fri,Golterman:2018mfm}. 
Going further away from the conformal window, we expect the dilaton state to merge into the lightest scalar state of the theory loosing its conformal properties, as properly encoded in the agnostic effective approach of~\cite{Hansen:2016fri}. This interest in the dilaton state is due to lattice studies of \(SU(3)\) gauge theories with matter field content consisting of $N_f= 8$ fundamental Dirac fermions~\cite{Appelquist:2016viq,Appelquist:2018yqe,Aoki:2014oha,Aoki:2016wnc}, and $N_f = 3$ symmetric 2-index Dirac fermions (sextet)~\cite{Fodor:2012ty,Fodor:2017nlp,Fodor:2019vmw} known as Minimal Walking Technicolor~\cite{Hong:2004td,Sannino:2004qp,Dietrich:2005jn,Evans:2005pu}. These studies reported evidence of the presence of a light singlet scalar particle in the spectrum, at least in the accessible range of fermion masses. It is therefore relevant to devise  independent tests of near-conformal dynamics. 

To gain novel information about  quantum phase transitions we propose, in this letter, to employ and extend the large-charge limit together with the state-operator correspondence~\cite{Cardy:1984rp,Cardy:1985lth,Hellerman:2015nra} which we assume to be approximately valid near the conformal fixed point. To elucidate our point we consider, for simplicity, a conformal theory with a \(U(1)\) global symmetry. We assume that the underlying dynamics is such that a quantum phase transition occurs breaking conformal invariance. To non-linearly realize the breaking of conformal invariance we include a dilaton in the \ac{eft}. We then restrict our attention to sectors of fixed \(U(1)\) charge \(Q\), which allows us to write a consistent \ac{eft} in the limit of \(Q \gg 1\).
The dilaton is, de facto, the  \emph{radial} mode of the \ac{eft}.  We determine the two-point function of the lowest operators in each charge sector. We discover that the mass of the dilaton which parametrises and quantifies the conformal symmetry breaking, induces a new logarithmic term in the charge. Thereby, analysing the large-charge limit we  gain novel relations aimed at isolating a signature of non-perturbative dilaton dynamics.  The generalization to non-Abelian global symmetry groups of immediate relevance for lattice investigations is straightforward and can be performed following~\cite{Lenaghan:2001sd}. 

\section{The dilaton as the radial mode}%
\label{sec:dilaton-as-radial}

We start our investigation by considering first an underlying \ac{cft} at large charge. In this limit 
 one considers sectors of fixed charge within a theory with a global symmetry.
In each sector the symmetry is spontaneously broken and the physics is described in terms of Goldstone bosons.
Even though the full theory is generically strongly coupled and cannot be accessed perturbatively, these Goldstone bosons can be described by an effective action.
In the limit where the charge is large, the semiclassical ground state dominates over the quantum fluctuations which are generically suppressed in inverse powers of the charge.
Since we work at fixed charge (as opposed to fixed charge density), we consider the system on \(\setR \times M_3\), where \(M_3\) is compact, in order to have an \ac{eft} that is well-defined everywhere.
As we will see later, we can however map our results to \(\setR^4\) using a Weyl rescaling.

Consider the simplest case of a \ac{cft} with a \(U(1)\) global symmetry.
Fixing the corresponding charge \(Q\) breaks this symmetry spontaneously and leads to a Goldstone boson.
We can describe its dynamics using a scale-invariant four-derivative action~\cite{Son:2002zn,Hellerman:2015nra,Monin:2016jmo}:
\begin{equation}
  \label{eq:NLSM}
  L_{NLSM}[\chi] = k_4 \pqty{ \del_\mu \chi \del_\mu \chi}^2 ,
\end{equation}
where \(k_4\) is an unknown coefficient that cannot be determined within the \ac{eft}~\footnote{This coefficient can be computed independently for example on the lattice~\cite{Banerjee:2017fcx,Banerjee:2019jpw}, at large-N~\cite{Alvarez-Gaume:2019biu}, or $4-\epsilon$~\cite{Watanabe:2019pdh,Badel:2019oxl}.}. There is also in principle a Wess--Zumino term that however contributes at lower order in \(1/Q\) and contains logarithmic corrections that vanish both on flat space and on the cylinder~\cite{Luty:2012ww,Hellerman:2015nra}.
The \(U(1)\) symmetry acts on \(\chi\) as \(\chi \to \chi + \delta\).
For simplicity, we will consider the theory on a torus of side \(L\), \(M_3 = T^3(L)\).
The classical solution at fixed charge \(Q\) is \(\chi = \mu t \), where \(\mu = (4 k_4 Q)^{1/3}/L \).
This solution spontaneously breaks the \(U(1)\) and leads to a Goldstone field \(\hat \chi\) whose action is obtained expanding the field in the \ac{nlsm} as \(\chi = \mu t + \hat \chi\).

This approach can be used more generally. We can start with a  two-derivative \ac{eft} for the prospective Goldstone of the type
\begin{equation}
  L_2[\chi] = \frac{f_\pi^2}{2} \del_\mu \chi \del_\mu \chi -  C^4,
\end{equation}
where \(f_\pi\) and \(C\) are dimension-one constants related to the underlying theory.
If we want to describe a (near) conformal theory,  we can introduce 
a new field \(\sigma\) -- the dilaton -- that under dilatations \(x \to e^\alpha x\) transforms as \(\sigma \to \sigma - \alpha/f\), where \(f\) is a constant of dimension \([f] = -1\). 
Using this field we can turn any action into a non-linearly realised conformally-invariant one by dressing all the operators \(\Op_k\) of dimension \([\Op_k] = k\) as
\begin{equation}
  \Op_k \to e^{(k-4) f \sigma} \Op_k .
\end{equation}
In our \(U(1)\) case we obtain
\begin{widetext}
\begin{equation}
  L_{CFT}[\chi,\sigma] = \frac{1}{2} g^{\mu\nu} f_\pi^{2} e^{-2\sigma f} \del_\mu \chi \del_\nu \chi - C^4 e^{- 4 \sigma f} + \frac{1}{2}e^{-2\sigma f}\pqty{g^{\mu\nu} \del_\mu\sigma \del_\nu \sigma- \frac{\xi R}{f^2}} +\order{R^2},
\end{equation}
\end{widetext}
where we have also added a kinetic term for the dilaton~\footnote{The dilaton here should not be confused with a modulus of the \ac{cft}. In the parlance of~\cite{Nicolis:2012vf} it is a gapped Goldstone that realizes approximate conformal invariance in the \ac{eft}. \acp{cft} with moduli space are typically supersymmetric and are described at large charge by a qualitatively different \ac{eft}~\cite{Hellerman:2017sur,Hellerman:2018xpi}.}. In view of wanting to invoke the state-operator correspondence, we have also added the Ricci scalar \(R\), the conformal coupling \(\xi = 1/6\), and the \(\order{R^2}\) terms that do not depend on the fields.
We now have obtained an effective action for the two Goldstones resulting from the breaking of the internal and of the conformal symmetry. 
From this point of view, the four-derivative action in Eq.~\eqref{eq:NLSM} can be viewed as the heavy-dilaton limit of this model.
The two fields can be combined into a complex dilaton, akin to the string-theoretical axio-dilaton:
\begin{equation}
  \Sigma = \sigma + i f_\pi \chi.
\end{equation}
Now the action can be recast in the form
\begin{equation}
  \label{eq:phi-4}
  L[\varphi] = \del_\mu \varphi^*\del^\mu\varphi - \xi R \varphi^* \varphi - u (\varphi^* \varphi)^2 + \order{R^2},
\end{equation}
where \(\varphi = 1/(\sqrt{2}f) e^{-f \Sigma}\), which means the dilaton appears as the radial mode of $\varphi$.
We are describing a \ac{cft}, which by definition has no dimensionful parameters.
The three dimensionful constants \(f_\pi\), \(C\) and \(f\) are combined into the two dimensionless quantities \(b = f f_\pi\) and \(u = 4 C^4 f^4\).
The former controls the deficit angle for the field \(\varphi\), which covers the whole complex plane only if \(b = 1\).
The action \(L[\varphi]\) was originally introduced in~\cite{Hellerman:2015nra,Alvarez-Gaume:2016vff} to describe the large-charge limit of the \(O(N)\) model.
The fixed-charge ground state is homogeneous and of the form
\begin{align}
  \chi &= \mu t, & \sigma &= \frac{1}{f} \log(v),   
\end{align}
where (on the torus) \(	\mu = 4 c_{4/3} \Lambda_Q /3 \), and \( v = 2 f_\pi \sqrt{c_{4/3}/3}/\Lambda_Q  \), and it has energy \(E = c_{4/3} Q^{4/3}/L\) where \(c_{4/3} = 3\pqty{C/(2 f_\pi)}^{4/3}\) and \(\Lambda_Q=Q^{1/3}/L\) is the scale associated to the fixed charge.

Expanding the fields around this \ac{vev} as \(\chi = \mu t + \hat \chi\) and \(\sigma = 1/f \log(v) + \hat \sigma\), and computing the propagator for the fluctuations \(\hat \chi\) and \(\hat \sigma\) we find one massless and one massive mode, with leading order dispersion relations  
\begin{align}
  \omega &= \frac{p}{\sqrt{3}} ,&  \omega &= b c_{4/3} \sqrt{\frac{32}{3} } \Lambda_Q
                                            + \frac{5 }{8 \sqrt{6}b c_{4/3} \Lambda_Q}  p^2.
\end{align}
The former is the expected conformal Goldstone that appears in all \acp{cft} at fixed charge.
The latter is a massive mode related to the dilaton.
This mode would have not appeared if we had not added a kinetic term for \(\sigma\) in the action and had used \(\sigma\) as a Lagrange multiplier.
The value of the classical energy would have however remained unchanged, since it is evaluated for constant \(\sigma\). The underlying non-perturbative information is efficiently parametrised by the two dimensionless parameters $b$ and $c_{4/3}$.

\section{A mass for the dilaton}%
\label{sec:dilaton-mass}

The construction in the previous section clarifies the role of the \emph{radial mode} of the field \(\varphi\) in the \(\varphi^4\) action in Eq.~\eqref{eq:phi-4}, but in the limit of large charge, where the massive mode decouples, it does not yet add more information on the \ac{cft} at hand.  It plays, however, a crucial role when extending the formulation to near-conformal theories.  This can be achieved by 
 adding a mass term for the dilaton \(\sigma \)~\cite{Coleman:1988aos}:
\begin{equation}
  \label{eq:walking-action}
  L_{m}[\chi, \sigma ] = L_{CFT}[\chi, \sigma] - U_m(\sigma),
\end{equation}
where
\begin{equation}
  U_m(\sigma) = \frac{m^2_\sigma}{16 f^2}\pqty{e^{-4 \sigma f} + 4 \sigma f -1}.	
\end{equation}
Here, \(m_\sigma\) is the mass of \(\sigma\) due to the underlying near-conformal dynamics. In fact, now the energy-momentum tensor is no longer traceless, its trace is proportional to the dilaton,
\begin{equation}%
  \label{eq:trace}
  T^\mu_{\phantom{\mu} \mu} = \frac{m_\sigma^2}{f} \sigma  . 
\end{equation}
It is through this operator that one encodes the (continuous) breaking of the conformal phase. For example, in perturbative  models of conformal symmetry breaking one can demonstrate that this is indeed the right operator, as can be seen from equation (13) and (15)  of~\cite{Antipin:2011aa}. In the non-perturbative regime, $m_\sigma$ still measures the amount of near-conformal dynamics for it is proportional, in gauge-fermion theories, to the beta function of the theory as explained in section VII.b of~\cite{Dietrich:2005jn} and in~\cite{Appelquist:2010gy}. 

Interestingly, the mass term \(U_m(\sigma)\) has a characteristic signature in the large-charge expansion of the physical observables. This is a welcome feature as it provides an independent handle when trying to disentangle the dilaton physics and features both analytically and via first-principle numerical simulations.

The first observation is that the near-conformal (walking) action in Eq.~\eqref{eq:walking-action} admits again a homogeneous fixed-charge solution of the same type as before.
On the torus we find that its energy is given by
\begin{equation}
  E = c_{4/3} \frac{Q^{4/3}}{L} - \frac{m_\sigma^2 L^3}{12 f^2}  \log(Q) + c_0 \ ,
\end{equation}
where \(c_0\) is a \(Q\)-independent constant.
As before, this result receives quantum corrections that are suppressed by powers of \(1/Q\).
The non-vanishing mass \(m_\sigma\) leads to a characteristic novel logarithmic term.
Expanding the fields around the ground state, we find again one massless and one massive mode:
\begin{align}
  \omega &= \frac{1}{\sqrt{3}} \pqty{ 1 + \frac{m_\sigma^2}{9  c_{4/3}  f^2\Lambda_Q^4}  } p , \\
  \omega &= b c_{4/3} \sqrt{\frac{32}{3} } \Lambda_Q + \frac{5 }{8 \sqrt{6}b c_{4/3} \Lambda_Q} \pqty{ 1 - \frac{m_\sigma^2}{20 c_{4/3} f^2 \Lambda_Q^4  }} p^2.
\end{align}
There is now, however, a contribution proportional to \(m_\sigma^2\) to the velocity of the putative conformal mode and a correction to the dispersion relation of the massive state as well. 

\bigskip

Another physical observable where the logarithmic behavior occurs is the conformal dimension of the lowest operator of charge \(Q\).
Strictly speaking, the conformal dimension is not defined in a non-conformal theory, but if we are sufficiently close to the fixed point and in a stationary point of the beta function of the full theory, the physics is still governed by the fixed point.
This means that under a Weyl rescaling of the metric \(g_{\mu\nu} \to \Omega(x) g_{\mu\nu} = g'_{\mu\nu}\), the operators in the theory transform as \(\Op(x) \mapsto \Omega(x)^{\Delta^*} \Op(x) = \Op'(x)\), were \(\Delta^*\) is the dimension in the reference \ac{cft} (\(m_\sigma = 0\)).
After analytic continuation, we can use the state-operator correspondence to compute two-point functions, mapping \(\setR^4\) to the cylinder frame.
Consider the Weyl transformation
\begin{align}
  g_{\mu\nu} &= \delta_{\mu\nu} \to \Omega(x) \delta_{\mu\nu} = g'_{\mu\nu} & \text{where \(\Omega(x) = r_0^2/\abs{x}^{2}\)}.
\end{align}
The metric \(g'\) describes a cylinder
\begin{equation}
  (\dd{s'})^2 = \dd{t}^2 + r_0^2 \dd{\Omega_3^2},
\end{equation}
where \(\abs{x} = r_0 e^{t/r_0}\) and \(\dd{\Omega_3^2}\) is the metric of the unit three-sphere.
The two-point function for the lowest operator of charge \(Q\) is given by~\cite{Monin:2016jmo,Hellerman:2017sur,Hellerman:2018xpi,Arias-Tamargo:2019xld,Badel:2019oxl}
\begin{widetext}
  \begin{equation}
    \ev{\Op_Q(t_0, \mathbf{n}_0) \Op_{-Q}(t_1, \mathbf{n}_1)}_{\text{cyl}} = \int \mathcal{D}\chi \mathcal{D}\sigma \exp[  Q \log(\varphi(t_0,\mathbf{n}_0) \bar{\varphi}(t_1, \mathbf{n}_1)) - \int \dd{t} \dd{\Omega} L_{m}[\chi, \sigma]],
  \end{equation}
\end{widetext}
where the first term describes the two insertions of an operator of charge \(Q\).
For large charge \(Q\), the path integral is dominated by the homogeneous saddle point \(\chi = i \mu t\), \(\sigma = \text{const.}\), 
\begin{equation}
  \ev{\Op_Q(t_0, \mathbf{n}_0) \Op_{-Q}(t_1, \mathbf{n}_1)}_{\text{cyl}} \approx e^{- E_{\text{cyl}} \abs{t_1 - t_0}},
\end{equation}
where \(E_{\text{cyl}}\) is the energy of the fixed-charge ground state on the cylinder
\begin{equation}
  r_0 E_{\text{cyl}} = \frac{c_{4/3}}{(4 \pi^2)^{1/3}}Q^{4/3} + c_{2/3} Q^{2/3} + c_0
  - \frac{\pi^2  m_\sigma^2 r_0^4}{3 f^2}  \log Q + \dots .
\end{equation}
and \(c_{2/3} = (\pi/(f_\pi \Lambda^2))^{2/3}/(2f^2)\).
For \(m_\sigma = 0\) this reproduces the expected behavior of a four-dimensional \ac{cft}~\cite{Orlando:2019hte}.
We can now map this expression to the two-point function in the flat-space frame,
\begin{multline}
  \ev{\Op_Q(x_0) \Op_{-Q}(x_1)}_{\text{flat}} \\ = e^{-\Delta^* (t_0+t_1)/r_0}\ev{\Op_Q(t_0, \mathbf{n}_0) \Op_{-Q}(t_1, \mathbf{n}_1)}_{\text{cyl}},
\end{multline}
where \(\Delta^*\) is the conformal dimensions in the reference \ac{cft}, which is given by the energy on the cylinder in the \(m_\sigma = 0\) limit \(\Delta^* = \eval{r_0 E_{cyl}}_{m_\sigma = 0} \).
Using translation invariance we can set \(t_0 \to -\infty\), \emph{i.e.} \(x_0 = 0 \) and we find the final result:
\begin{equation}
  \ev{\Op_Q(0) \Op_{-Q}(x)}_{\text{flat}} = \frac{c_Q}{\abs{x}^{\Delta^* + r_0 E_{\text{cyl}}}} = \frac{c_Q}{\abs{x}^{2\Delta}} ,
\end{equation}
where \(c_Q\) is a normalization constant, and
\begin{equation}
  \Delta = \Delta^* \pqty{ 1 - \frac{m_\sigma^2}{24 c_{4/3} f^2 \Lambda_Q^4} \log Q + \dots}.
\end{equation}
As  observed in~\cite{Banerjee:2017fcx,Banerjee:2019jpw}, the leading coefficients in the large-charge expansion of the energy on the torus and in the conformal dimension are related via the \ac{eft} even though the cylinder and the torus are not conformally equivalent.
Once more, we find the characteristic logarithmic term in the charge that marks the departure from conformal invariance.
The latter appears naturally because the underlying dynamics generates a new scale, whose square is $m_\sigma/f$, that contributes to the trace of the energy-momentum tensor in Eq.~\eqref{eq:trace}. 

\section{Conclusions}
The large-charge limit has been adapted and extended to study four-dimensional near-conformal dynamics.   We enforce the latter  by augmenting the low-energy theory with a dilaton which, in large-charge parlance, is  related to the \emph{radial} mode of the \ac{eft}.
We compute the ground-state energy in sectors of large charge on the torus and the two-point function of charged operators on the cylinder. 
The presence of (near) conformal dynamics permits to use the state-operator correspondence and derive the two-point function in flat space.
We find that the mass of the dilaton induces a novel term, logarithmic in the charge.
This shows that the large-charge limit provides a new handle to explore near-conformal dynamics while testing dilaton-related  properties.
The approach can be readily extended to other space-time dimensions and non-Abelian global symmetry groups.

\subsection*{Acknowledgments}

D.O. and S.R. would like to thank Luis Álvarez-Gaumé and Simeon Hellerman for useful discussions.
The work of S.R. is supported by the Swiss National Science Foundation under grant number {PP00P2\_183718/1}.
The work of F.S. is partially supported by the Danish National Research Foundation grant DNRF:90. 
D.O. acknowledges partial support by the {NCCR 51NF40--141869} ``The Mathematics of Physics'' (Swiss\textsc{map}).
We would like to thank the Simons Center for Geometry and Physics for  hospitality during the final phases of this work.

\bibliography{References}

\end{document}